\documentclass[aps,prl,twocolumn,superscriptaddress,showpacs,showkeys,amssymb,amsmath]{revtex4-1}
\usepackage{graphicx}

\newcommand{\V}[1]{\mathbf{#1}} 
\newcommand{\T}[1]{\texttt{#1}} 
\newcommand\Alfven{Alfv\'en }

\newcommand{\figref}[1]{Fig.~\ref{#1}}

\begin{document}

\title{Gyrokinetic Simulations of Solar Wind Turbulence from Ion to Electron Scales}

\author{G.~G. Howes}
\email[]{gregory-howes@uiowa.edu}
\affiliation{Department of Physics and Astronomy, University of Iowa, Iowa City, 
Iowa 52242, USA.}
\affiliation{Isaac Newton Institute for Mathematical Sciences, Cambridge, CB3 0EH, U.K.}

\author{J.~M. TenBarge}
\affiliation{Department of Physics and Astronomy, University of Iowa, Iowa City, 
Iowa 52242, USA.}

\author{W.~Dorland}
\affiliation{Department of Physics, University of Maryland, College Park, 
Maryland 20742-3511, USA.}
\affiliation{Isaac Newton Institute for Mathematical Sciences, Cambridge, CB3 0EH, U.K.}

\author{E.~Quataert}
\affiliation{Department of Astronomy, University of California, Berkeley, 
California 94720, USA.}

\author{A.~A.~Schekochihin}
\affiliation{Rudolf Peierls Centre for Theoretical Physics, University of Oxford, Oxford, OX1 3NP, U.K.}
\affiliation{Isaac Newton Institute for Mathematical Sciences, Cambridge, CB3 0EH, U.K.}

\author{R.~Numata}
\affiliation{Department of Physics, University of Maryland, College Park, 
Maryland 20742-3511, USA.}
\affiliation{Isaac Newton Institute for Mathematical Sciences, Cambridge, CB3 0EH, U.K.}

\author{T.~Tatsuno}
\affiliation{Department of Physics, University of Maryland, College Park, 
Maryland 20742-3511, USA.}

\date{\today}

\begin{abstract}
The first three-dimensional, nonlinear gyrokinetic simulation of
plasma turbulence resolving scales from the ion to electron gyroradius
with a realistic mass ratio is presented, where all damping is
provided by resolved physical mechanisms.  The resulting energy
spectra are quantitatively consistent with a magnetic power spectrum
scaling of $k^{-2.8}$ as observed in \emph{in situ} spacecraft
measurements of the ``dissipation range'' of solar wind turbulence.
Despite the strongly nonlinear nature of the turbulence, the linear
kinetic \Alfven wave mode quantitatively describes the polarization of
the turbulent fluctuations. The collisional ion heating is measured at
sub-ion-Larmor radius scales, which provides the first evidence of the
ion entropy cascade in an electromagnetic turbulence simulation.
\end{abstract}


\maketitle 

\emph{Introduction.}---Although studies of the inertial range of 
magnetized plasma turbulence date back more than four decades
\cite{Coleman:1968}, only recently has  the ``dissipation
range''---the kinetic scales from the ion to the electron Larmor
radius and beyond---become a focus of the astrophysics and
heliospheric physics communities. The dynamics in the dissipation
range is critical to a fundamental understanding of plasma turbulence
because it is at these scales that the turbulence is dissipated and
the turbulent energy is converted into ion and electron thermal
energy.
One of the principal challenges is that the dynamics at these scales is
weakly collisional in most diffuse space and astrophysical plasmas.
Therefore, a kinetic description of the plasma dynamics and energy
conversion via wave-particle interactions is necessary
\cite{Marsch:2006,Howes:2008b,Schekochihin:2009}. This is
significantly more challenging than the fluid descriptions, such as
magnetohydrodynamics (MHD), commonly used in the study of the
inertial-range turbulence. To enable direct comparisons of numerical
results to observations of the dissipation range, one must cover a
meaningful dynamic range while satisfying three important conditions:
the turbulence must be modeled in three spatial dimensions
\footnote{Three spatial dimensions are required because the dominant
$\V{E}\times\V{B}$ nonlinearity cannot exist unless both dimensions
perpendicular to the mean field are included, while Alfv\'enic wave
propagation requires the parallel dimension as well
\cite{Howes:2006,Schekochihin:2009}}, kinetic dissipation mechanisms must be resolved, and a
realistic mass ratio must be used. 


Early observational studies of the solar wind probed little more than
a decade in satellite-frame frequency above the spectral break marking
the onset of the dissipation range \cite{Leamon:1999}. More recent,
better resolved measurements show that the one-dimensional magnetic
energy spectrum typically exhibits a power-law behavior with a $-5/3$
spectral index at low frequencies, a break at a few tenths of a Hz,
and a steeper apparent power-law spectrum at higher frequencies. The
spectral exponent in this high-frequency range reported in most recent
studies ranges from $-2.6$ to $-2.8$
\cite{Sahraoui:2009,Kiyani:2009,Alexandrova:2009,Chen:2010,Sahraoui:2010b}.
Ideas proposed to explain the steeper dissipation range spectrum
include proton cyclotron damping \cite{Coleman:1968}, Landau damping
of kinetic \Alfven waves \cite{Leamon:1999,Howes:2008b}, and the dispersion of whistler 
waves \cite{Stawicki:2001}.  

Following modern theories of anisotropic MHD turbulence
\cite{Goldreich:1995}, a theoretical picture of the kinetic turbulent
cascade was developed \cite{Quataert:1999,Howes:2008c,Schekochihin:2009} that
proposed an inertial range containing an anisotropic cascade of MHD
\Alfven waves at $k_\perp \rho_i \ll 1$, a break in the magnetic 
energy spectrum at the perpendicular scale of the ion Larmor radius
$k_\perp \rho_i \sim 1$, and a dissipation range of kinetic \Alfven
waves (KAWs) at $k_\perp \rho_i \gg 1$.  Because of the spatial
anisotropy of the fluctuations ($k_\perp \gg k_\parallel$), the
turbulent fluctuation frequencies stay well below the ion cyclotron
frequency, $\omega \ll \Omega_i$, in both the inertial and dissipation
ranges.  At scales $k_\perp \rho_i \gtrsim 1$, collisionless damping
via the Landau resonance with ions and electrons dissipates the
turbulence. Numerical evidence that this picture yields
observationally sensible predictions was provided by the first fully
electromagnetic, 3D, kinetic simulations of turbulence over the range $0.4
\le k_\perp\rho_i \le 8$
\cite{Howes:2008a} using the gyrokinetic equations
\cite{Frieman:1982,Howes:2006,Schekochihin:2009}, which are a rigorous
low-frequency anisotropic limit of kinetic theory.

Over the past two years, a number of observational studies have
broken exciting new ground by probing the dynamics of solar wind
turbulence up to satellite-frame frequencies $f \sim 100$~Hz
\cite{Sahraoui:2009,Kiyani:2009,Alexandrova:2009,Chen:2010,Sahraoui:2010b}.  
All of these observations clearly demonstrate a nearly power-law
behavior of the magnetic power spectrum over the dissipation range, $1
\mbox{ Hz} \lesssim f \lesssim 50 \mbox{ Hz}$. This finding is in 
disagreement with cascade models of KAW turbulence that employed
linear Landau damping rates and predicted that the latter would be
sufficient to lead to an exponential fall-off of the spectrum before
it reached the electron scales \cite{Howes:2008b,Podesta:2010a}. This
raises the important question of whether the KAW cascade can explain
the observed spectra. 
Other measurements are consistent with the key assumptions of KAW
turbulence models: the turbulence remains anisotropic on these scales
\cite{Podesta:2009a,Chen:2010,Sahraoui:2010b}; and the fluctuations
appear consistent with a KAW-like polarization
\cite{Sahraoui:2009,Sahraoui:2010b} (see, however, \cite{Chen:2010}).


This Letter presents the first nonlinear gyrokinetic simulation of
solar wind turbulence resolving the entire range of scales from the
ion (proton) to the electron Larmor radius with the correct mass
ratio. The significant advance achieved by this simulation is that no
artificial dissipation is required to remove energy at small scales;
all dissipation is due to resolved collisionless damping via the
Landau resonances.  The steady-state spectra may therefore be compared
directly to observations of the solar wind dissipation range
\footnote{A Maxwellian equilibrium distribution is assumed, so effects
such as the generation of the firehose and mirror instabilities by
temperature anisotropies \cite{Bale:2009} are not included}. The
resulting numerical spectra are quantitatively consistent with the
recent \emph{in situ} satellite observations
\cite{Sahraoui:2009,Kiyani:2009,Alexandrova:2009,Sahraoui:2010b}, demonstrating that
a KAW cascade, even when the physical damping mechanisms are taken
into account, can produce a nearly power-law behavior over the
dissipation range.

\emph{Simulation.}---The simulation in this paper was performed 
using \T{AstroGK}, the Astrophysical Gyrokinetics Code, developed
specifically to study kinetic turbulence in astrophysical plasmas.  A
detailed description of the code and the results of linear and
nonlinear benchmarks are presented in \cite{Numata:2010}, so we give
here only a brief overview.

\T{AstroGK} evolves the perturbed gyroaveraged distribution
function $h_s(x,y,z,\lambda,\varepsilon)$ for each species $s$, the
scalar potential $\varphi$, parallel vector potential $A_\parallel$,
and the parallel magnetic field perturbation $\delta B_\parallel$
according to the gyrokinetic equation and the gyroaveraged Maxwell's
equations \cite{Frieman:1982,Howes:2006}. The velocity space
coordinates are $\lambda=v_\perp^2/v^2$ and $\varepsilon=v^2/2$. The
domain is a periodic box of size $L_{\perp }^2 \times
L_{\parallel }$, elongated along the straight, uniform mean magnetic
field $B_0$. Note that, in the gyrokinetic formalism, all quantities may
be rescaled to any parallel dimension satisfying $L_{\parallel }
/L_{\perp } \gg 1$. Uniform Maxwellian equilibria for ions (protons)
and electrons are chosen, and the correct mass ratio $m_i/m_e=1836$ is
used. Spatial dimensions $(x,y)$ perpendicular to the mean field are
treated pseudospectrally; an upwind finite-difference scheme is used
in the parallel direction, $z$. Collisions are incorporated using a
fully conservative, linearized collision operator that includes
energy diffusion and pitch-angle scattering
\cite{Abel:2008,Barnes:2009}.

Because turbulence in astrophysical plasmas exists over a wide
range of scales---for example, in the near-Earth solar wind, the driving scale
is $L \sim 10^{6}$~km and the electron Larmor radius is $\rho_e \sim
1$~km \cite{Howes:2008b}---numerical simulations are
necessarily limited to modeling only a portion of the turbulent
cascade. Two key challenges in the kinetic simulation of turbulence
are modeling the energy injection at the largest scales and removing
the turbulent energy at the smallest resolved scales.

In gyrokinetic simulations using \T{AstroGK}, the simulation domain is
much smaller than the physical driving scale, and covers scales at
which the turbulent fluctuations are sufficiently anisotropic ($k_\perp
\gg k_\parallel$) for the gyrokinetic approximation to be well
satisfied
\cite{Howes:2006,Howes:2008b,Schekochihin:2009}; this assumption of anisotropy 
is consistent with recent multi-point spacecraft measurements
\cite{Chen:2010,Sahraoui:2010b}.  In the simulation reported here,
nonlinear energy transfer from turbulent fluctuations at scales larger
than the largest scales in our domain ($k_{\perp 0} \rho_i=1$) is
modeled using six modes of a parallel ``antenna'' current
$j_{\parallel,\mathbf{k}}^a$ added via Amp\`ere's Law
\cite{Numata:2010}.  These driven modes have wavevectors $(k_x
\rho_i,k_y \rho_i,k_z L_\parallel/2 \pi)=(1,0,\pm 1), (0,1,\pm 1),
(-1,0,\pm 1)$, frequencies $\omega_a = 1.14 \omega_{A0}$ (where
$\omega_{A0}
\equiv k_{\parallel 0}v_A$ is a characteristic \Alfven frequency corresponding 
to the parallel size $L_\parallel$ of the domain), and amplitudes that evolve
according to a Langevin equation. This produces Alfv\'enic wave modes
with a frequency $\omega
\sim \pm k_{\parallel 0} v_A$ and a decorrelation rate comparable to
$\omega$, as expected for critically balanced Alfv\'enic turbulence
\cite{Goldreich:1995}. Note that energy is injected only at $k_\perp
\rho_i=1$, so the amplitudes at all higher  $k_\perp$  are determined
by the  nonlinear dynamics.

The key advance achieved here is resolving a fully dealiased range
with a perpendicular scale separation of $\sqrt{m_i/m_e} \simeq
42.8$. This spans from the proton Larmor radius at $k_\perp \rho_i=1$ to
the electron Larmor radius at $k_\perp \rho_e=1$ (or $k_\perp \rho_i
\simeq 42.8$ for $T_i/T_e=1$). Wave-particle interactions
via the Landau resonance are resolved and sufficient to damp the
electromagnetic fluctuations within the simulated range of scales.
The energy spectra at all scales, including the dissipative scales,
are shaped by resolved physical processes, and so may be
directly compared to observational data. This would not be possible
with standard fluid or hybrid approaches, such as Hall MHD or kinetic
ion-fluid electron models, because the bulk of the damping of the
electron motion in such models is typically governed by an
\emph{ad hoc} fluid model of the damping, such as viscosity or
resistivity \cite{Howes:2008d}.

The simulation domain size is $L_{\perp}=2\pi \rho_i$ with dimensions
$(n_x,n_y,n_z,n_\lambda,n_\varepsilon,n_s)= (128,128,128,64,16,2)$. The plasma
parameters are $\beta_i=1$ and $T_i/T_e=1$. In kinetic turbulence
simulations, the inclusion of sufficient collisionality is essential
to prevent the small-scale structure in velocity space generated by
wave-particle interactions from exceeding the velocity space
resolution. For this reason, collision frequencies of $\nu_i=0.04
\omega_{A0}$ and $\nu_e=0.5 \omega_{A0}$ have been chosen carefully to
achieve sufficient damping of small-scale velocity-space structure yet
to avoid altering the collisionless dynamics of each species over the
range of scales at which the kinetic damping is non-negligible.  A
recursive expansion procedure \cite{Howes:2008c} is used to reach a
statistically steady state at acceptable numerical cost. At low
spatial resolution, the simulation is run for more than an outer-scale
eddy turnover time (this turnover time is  $\tau_0=5.52
\omega_{A0}^{-1}$) to reach a steady state. Resolution in each spatial dimension 
is then doubled, and the simulation is run to a new steady state,
which requires only a time of order the cascade time at the smallest
mode before expansion. This simulation has been evolved, using this
recursive procedure, to a time of $t=18.70 \omega_{A0}^{-1}$.


\begin{figure}
\resizebox{3.0in}{!}{\includegraphics*[0.65in,2.in][8.1in,7.8in]{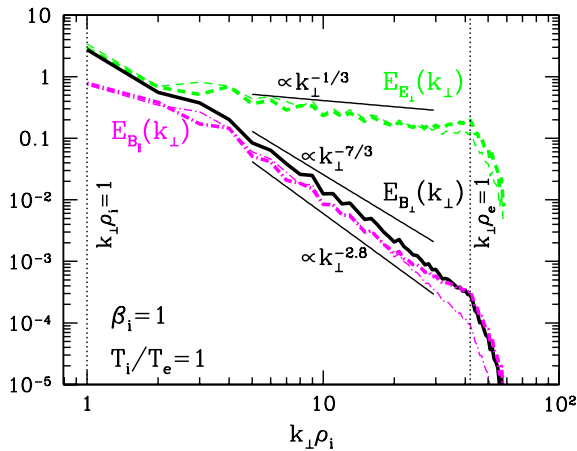}}
\caption{ \label{fig:milestone}  For a $\beta_i=1$ and
$T_i/T_e=1$ plasma, thick lines are numerical energy spectra for the
perpendicular magnetic (solid), electric (dashed), and parallel
magnetic (dot-dashed) field perturbations.  Thin lines are the
perpendicular electric (dashed) and parallel magnetic (dot-dashed)
energy spectra predicted from the perpendicular magnetic energy
spectrum in the simulation assuming the polarization of a linear
collisionless KAW---the result is in excellent agreement with the
simulation for $k_\perp \rho_i \lesssim 30$. Ion and electron Larmor
radius scales are marked by vertical dotted lines.  }
\end{figure}

\emph{Spectra.}---In \figref{fig:milestone}, steady-state 
energy spectra (thick lines) are presented, including the
perpendicular magnetic (solid), parallel magnetic (dot-dashed), and
perpendicular electric (dashed) energy spectra. Normalizations of the
energy spectra are the same as in our previous work
\cite{Howes:2008a}. The salient feature of the perpendicular magnetic
energy spectrum is its nearly power-law appearance over the entire
range of scales from $k_\perp \rho_i=1$ to $k_\perp \rho_e=1$.  For a
KAW cascade, the theoretical prediction for its scaling in the absence
of dissipation is $k_\perp ^{-7/3}$
\cite{Howes:2008b,Schekochihin:2009}; the measured spectrum over the
range $5 \le k_\perp \rho_i \le 30$ is slightly steeper, about
$k_\perp^{-2.8}$, in remarkable agreement with recent observations
\cite{Kiyani:2009,Alexandrova:2009,Sahraoui:2010b}. This suggests that resolving the
damping via the Landau resonance with electrons in this region is
important to enable direct comparison to satellite observations.
Clearly, the energy spectra of this gyrokinetic simulation do not
exhibit an exponential fall-off. It appears that the effect of
(relatively weak) electron Landau damping is merely to steepen the
spectra slightly.  For a $\beta_i=1$ plasma, the only propagating wave
mode in gyrokinetics at these scales is the KAW, so this numerical
result disproves by counterexample a recent claim
\cite{Podesta:2010a} that the KAW cascade cannot reach scales of order
the electron gyroradius $k_\perp
\rho_e \sim 1$.

\begin{figure}
\resizebox{3.0in}{!}{\includegraphics*[0.3in,2.4in][8.05in,9.55in]{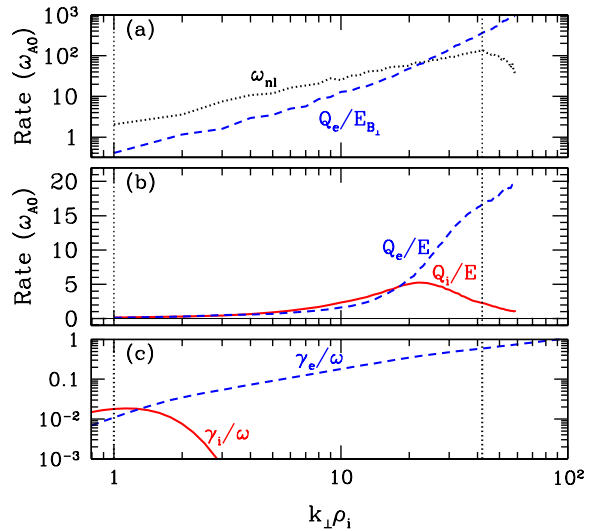}}
\caption{ \label{fig:heat} (a) ``Nonlinear damping rate'' 
$Q_e/E_{B_\perp}$ and nonlinear transfer frequency
$\omega_{nl}$ show strong damping for $k_\perp \rho_i \gtrsim 25$.
(b) Ion (solid) and electron (dashed) collisional heating
vs.~$k_\perp$, normalized to generalized energy $E$
\cite{Schekochihin:2009}. (c) The linear ion (solid) and 
electron (dashed) Landau damping rates. }
\end{figure}

\emph{KAW Polarized Turbulence.}---An important question in the study 
of plasma turbulence at small scales is whether linear wave properties
can provide useful guidance in exploring the characteristics of the
nonlinear fluctuations in a turbulent plasma. To investigate this
question, we show in \figref{fig:milestone} the spectra for
$E_{E_\perp}(k_\perp)$ (thin dashed) and $E_{\delta
B_\parallel}(k_\perp)$ (thin dot-dashed), predicted from the numerical
spectrum of $E_{B_\perp}(k_\perp)$ (thick solid) by assuming that the
relationships between the different components of the fields are
described by the linear eigenfunctions of the KAW derived from the
linear collisionless gyrokinetic dispersion relation
\cite{Howes:2006}. The predicted spectra are in excellent agreement
with the numerical spectra for $E_{E_\perp}(k_\perp)$ (thick dashed)
and $E_{\delta B_\parallel}(k_\perp)$ (thick dot-dashed) for scales
$k_\perp
\rho_i \lesssim 30$, giving strong support to the view that the linear
wave properties of the kinetic plasma are crucial in determining the
nature of the turbulent fluctuations. Note that recent measurements of
solar wind turbulence have found consistency with KAW polarization
\cite{Sahraoui:2010b}. Only for $k_\perp \rho_i \gtrsim 30$ does 
the nature of the fluctuations in our simulation deviate significantly
from that of the linear collisionless KAW; this may be due to the
effect of collisions (see \figref{fig:heat}).

We stress that while the relative polarization of the turbulent
fluctuations follows the linear theory, this does
\emph{not} mean that the turbulence is weak---in fact, the $k_\perp^{-7/3}$
prediction comes from assuming a critically balanced, and therefore
strong, KAW turbulence \cite{Howes:2008b,Schekochihin:2009}, which
means that the fluctuations decorrelate over a timescale comparable to
the linear wave period.

\emph{Heating.}---One of the fundamental goals in the study of the 
dissipation range of plasma turbulence is to identify the physical
mechanisms by which the energy in turbulent fluctuations is converted
into plasma heat. By Boltzmann's $H$ theorem, the increase of entropy
and associated irreversible heating in a kinetic plasma is possible
only by the action of collisions \cite{Howes:2006}. Collisionless
wave-particle interactions transfer electromagnetic fluctuation energy
to the perturbed particle distribution functions.  It has been
proposed that nonlinear phase mixing, due to the variation in the
gyroaveraged $\V{E} \times\V{B}$ drift velocity at sub-Larmor radius
scales, causes the non-thermal energy in the distribution function to
undergo a phase-space cascade to smaller scales in both physical and
velocity space, the so-called entropy cascade
\cite{Schekochihin:2009}.  The entropy cascade  ultimately
enables irreversible thermodynamic heating in a weakly collisional
plasma by driving the nonthermal energy to small scales in velocity
space, where even weak collisions can act effectively. The Landau
damping merely transfers energy from the electromagnetic fluctuations
to the entropy cascade.

The gyrokinetic heating equations \cite{Howes:2006} have been
implemented in \T{AstroGK} to diagnose the plasma collisional heating
rate by species, $Q_s$. Dissipation of the KAW cascade by electron
Landau damping in the simulation is sufficient to terminate the
cascade, demonstrated in
\figref{fig:heat} (a) by comparing the ``nonlinear damping
rate'' $\gamma_{nl}\sim Q_e/E_{B_\perp}$ (dashed) to the frequency of
the nonlinear energy transfer $\omega_{nl} \sim k_\perp v_A (\delta
B_\perp/B_0) (1+k_\perp^2\rho_i^2/2)^{1/2}$ \cite{Howes:2008b}
(dotted). Here $E_{B_\perp}$ is the perpendicular magnetic energy of
the KAW cascade, spectra are summed in linearly spaced bins in
$k_\perp$, and time is given in units of $\omega_{A0}^{-1}$.  A key
result of this Letter, in \figref{fig:heat} (b), is the ion (solid)
and electron (dashed) collisional heating rate normalized to the
generalized energy $E$ (which includes energy from both KAW and ion
entropy cascades)
\cite{Schekochihin:2009}. Also plotted (c) are the linear
collisionless Landau damping rates for ions (solid) and electrons
(dashed) in the gyrokinetic limit. Although the linear damping onto
the ions peaks at $k_\perp \rho_i \sim 1$, the peak of the ion
collisional heating occurs at the higher wavenumber of $k_\perp \rho_i
\sim 20$. This shift of the peak of ion collisional heating to
higher wavenumber is consistent with the predicted effect of the ion
entropy cascade, which transfers energy to sub-Larmor scales.
Our simulation results thus provide the first evidence of the ion
entropy cascade in a 3D, driven, electromagnetic turbulence
simulation, previously accomplished only for a 2D, decaying,
electrostatic case \cite{Tatsuno:2009}.

\emph{Conclusions.}---We have presented the first nonlinear kinetic 
simulation of solar wind turbulence resolving the entire range of
scales from the ion (proton) to the electron Larmor radius with the
correct mass ratio. No artificial dissipation was required to remove
energy at the small scales; all dissipation was due to resolved
collisionless damping mechanisms.  The resulting steady-state energy
spectra are quantitatively consistent with recent \emph{in situ}
satellite observations which demonstrate nearly power-law behavior
over the dissipation range. Although the turbulence is strong, the
relative polarizations of the electromagnetic fluctuations in this
nonlinearly turbulent system are still quantitatively described by the
linear kinetic \Alfven wave mode over much of the dissipation range.
Although the collisionless ion damping of the electromagnetic
fluctuations peaks at $k_\perp \rho_i \sim 1$, collisional ion heating
in the numerical simulation peaks at a much higher wavenumber of
$k_\perp \rho_i \sim 20$. This is consistent with the presence of the
theoretically predicted ion entropy cascade \cite{Schekochihin:2009},
responsible for mediating entropy increase and irreversible
thermodynamic heating in weakly collisional plasmas such as the solar
wind. Future numerical and analytical studies of the kinetic \Alfven
wave cascade and its dissipation will aim to
determine quantitatively the plasma heating as a
function of plasma parameters.

This work was supported by the DOE Center for Multiscale Plasma
Dynamics, STFC, Leverhulme Trust Network for Magnetised Plasma
Turbulence, NSF-DOE PHY-0812811, and NSF ATM-0752503.  Computing
resources were supplied through DOE INCITE 2008 Award PSS002, NSF
TeraGrid 2009 Award PHY090084, and DOE INCITE Award FUS030.


%

\end{document}